\documentclass[conference]{IEEEtran}

\usepackage{epsfig,subfigure}
\usepackage{amssymb, amsmath}
\usepackage{color}

\usepackage{graphicx}

\usepackage{amssymb}
\usepackage{amsmath,amsfonts,amssymb}
\usepackage{verbatim}
\usepackage[bookmarks=false]{}
\usepackage[font={small}]{caption}

\newtheorem{lemma}{Lemma}

\newtheorem{ex}{\textbf{Example}}

\begin{document}
\date{}
\title{Multilevel Topological Interference Management}
\author{  Chunhua Geng, Hua Sun, and Syed A. Jafar  \\
 Center for Pervasive Communications and Computing (CPCC)\\
 University of California Irvine, Irvine, CA, USA
}

\maketitle

\begin{abstract}
The robust principles of treating interference as noise (TIN) when it is sufficiently weak, and avoiding it when it is not, form the background for this work. Combining TIN with the topological interference management (TIM) framework that identifies optimal interference avoidance schemes, a  baseline TIM-TIN approach is proposed which decomposes a network into TIN and TIM components,  allocates the signal power levels  to each user in the TIN component, allocates signal vector space dimensions  to each user in the TIM component, and guarantees that the product of the two  is an achievable number of signal dimensions available to each user in the original network.
\end{abstract}

\section{Introduction}
The capacity of wireless interference networks is a rapidly evolving research front, spurred in part by exciting breakthroughs such as the idea of interference alignment \cite{Jafar_FnT} which provides fascinating theoretical insights and shows much promise  under idealized conditions. The connection to practical settings however remains tenuous. This is in part due to the following two factors. First,  because of the assumption of precise channel knowledge, idealized studies often get caught in the minutiae of channel realizations, e.g., rational versus irrational values, that have little bearing in practice. Second, by focusing on the degrees of freedom (DoF) of fully connected networks,  these studies ignore the most critical aspect of interference management in practice -- the differences of signal strengths due to path loss and fading (in short, network topology). Indeed, the DoF metric treats every channel as essentially equally strong (capable of carrying exactly 1 DoF). So the desired signal has to actively avoid \emph{every} interferer, whereas in practice each user needs to avoid only  a few significant interferers and the rest are weak enough to be safely ignored. Therefore, by trivializing the topology of the network, the DoF studies of fully connected networks make the problem much harder than it needs to be. Non-trivial solutions to this harder problem  invariably rely on much more channel knowledge than is available in practice. Thus, the two limiting factors re-enforce each other. 

Evidently, in order to avoid these pitfalls, one should shift focus away from optimal ways of exploiting precise channel knowledge (which is rarely available), and toward optimal ways of exploiting a coarse knowledge of interference network \emph{topology}. This line of thought motivates robust models of interference networks  where  only a {coarse} knowledge of channel strength {levels}  is available to the transmitters and no channel phase knowledge is assumed. This is the \emph{multilevel} topological interference management framework. It is a generalization of the elementary topological interference management framework introduced in \cite{Jafar_TIM}, wherein the transmitters can only distinguish between channels that are connected (strong) and not connected (weak). 




\section{Robust Principles of Interference Management: Ignore, Avoid}
Existing wireless interference networks are based on two robust interference management principles ---  1) \emph{ignore}  interference that is sufficiently weak, and 2) \emph{avoid}  interference that is not. In slightly more technical terms, ignoring interference translates into treating it as noise, and avoiding interference translates into access schemes such as TDMA/FDMA/CDMA. The intuitive appeal of these principles lies in their robustness, and in particular, their minimal channel knowledge requirements. Recent work has explored the optimality of both of these principles. 


\begin{enumerate}
\item {\it TIN:} The optimality of the first principle, \emph{treating interference as noise} (TIN) when it is sufficiently weak, has received much attention \cite{Annapureddy_Veeravalli_TIN_opt,Motahari_Khandani_TIN_opt,Kramer_TIN_opt, Annapureddy_Veeravalli_MIMO,Tse_GDoF,Jafar_Vishwanath_GDoF}. Most recently, in \cite{Geng_TIN_opt}, Geng et al. show that in a general $K$ user Gaussian interference channel setting, if for each user the desired signal strength is no less than the sum of the strengths of the strongest interference  \emph{from} this user and the strongest interference \emph{to} this user (all values in dB scale), then TIN is the optimal scheme for the entire capacity region of this network, up to a constant gap of no more than $\log(3K)$ bits.

\item {\it TIM:} The optimality of the second principle, \emph{avoidance}, has been investigated most recently by \cite{Jafar_TIM}, as the  \emph{topological interference management} (TIM) problem. With channel knowledge at the transmitters limited to a coarse knowledge of network topology (which links are stronger/weaker than the effective noise floor),  TIM  is shown in \cite{Jafar_TIM} to be essentially  an index coding problem \cite{Birk_Kol}.  TIM subsumes within itself the TDMA/FDMA/CDMA schemes as trivial special cases, but is in general much more capable  than these conventional approaches. 
\end{enumerate}

\section{TIM-TIN: Joint View of Signal Vector Spaces and Signal Power Levels}
The two principles -- avoiding versus ignoring interference -- which are mapped to TIM and TIN, respectively, naturally correspond to interference management in terms of signal  \emph{vector spaces} and signal \emph{power levels}. TIM uses the interference alignment perspective \cite{Jafar_TIM, Maleki_Cadambe_Jafar_index} to optimally allocate signal vector subspaces among the  interferers. Note that in order to resolve desired signal from interference based on the signal vector spaces, the strength of each signal is irrelevant. What matters is only that desired signal and the interference occupy linearly independent spaces. TIN, on the other hand, optimally allocates signal power levels among users by setting the transmit power  level at each transmitter and the noise floor level at each receiver. Thus TIN depends very much on the strengths of signals relative to each other. Associating TIM with signal vector space allocations and TIN with signal power level allocations within the multilevel topological interference management framework, we refer to the joint allocation of signal vector spaces and signal power levels as the TIM-TIN problem.


{\it TIM-TIN Problem:} With only a coarse knowledge of channel strengths available to the transmitters, we wish to carefully allocate not only the beamforming vector {directions} (signal vector spaces) but also the transmit powers (signal power levels) to each of those beamforming vectors. The necessity of a joint TIM-TIN perspective is evident as follows. In vector space allocation schemes used for DoF studies, the signal space containing the interference is entirely rejected (zero-forced). This is typically fine for linear DoF studies because all signals are essentially equally strong,  every substream carries one DoF, so any desired signal projected into the interference space cannot achieve a non-zero DoF. However, once we account for the difference in signal strengths in a generalized degrees of freedom (GDoF) framework,  the signal vector space dimensions occupied by interference may not be \emph{fully} occupied in terms of power levels if the interference is weak. So, non-zero GDoF may be achieved by desired signals projected into the same dimensions as occupied by the interference, where interference is weaker than desired signal. It is this aspect that we wish to exploit in the multilevel topological interference management framework. In this preliminary work, which represents our first steps in this direction, our focus will not be on optimality, but rather on simplicity and robustness.  In particular, we will formulate the problem and identify a natural baseline against which more sophisticated schemes  may be compared.

\section{Channel Model}

Similar to \cite{Geng_TIN_opt}, we represent the  channel model in the following  form,
\begin{equation}
\label{equ_channel}
\begin{aligned}
y_k(t)=\sum_{i=1}^K\sqrt{P^{\alpha_{ki}}}e^{j\theta_{ki}}x_i(t)+z_k(t),~~~ \forall k\in\{1,2,...,K\}.
\end{aligned}
\end{equation}
where at each time index $t$, $x_i(t)$ is the transmitted symbol of transmitter $i$, $y_k(t)$ is the received signal at receiver $k$, $\sqrt{P^{\alpha_{ki}}}$ and $\theta_{ki}$ are the complex channel gain and phase value from transmitter $i$ to receiver $k$, and $z_k(t)\sim \mathcal{CN}(0,1)$ is the additive white Gaussian noise (AWGN) at receiver $k$. All  symbols are complex. Each transmitter $i$ is subject to the power constraint $E[|x_i(t)|^2]\leq 1$. We will call the exponent $\alpha_{ki}$ the channel strength level of the link between transmitter $i$ and receiver $k$, and as done in \cite{Geng_TIN_opt}, we will assume the $\alpha_{ki}$ values are non-negative (the negative values are mapped to zero).   

The definitions of messages, achievable rates, capacity region, Generalized degrees of freedom (GDoF) are all  standard  (see, e.g., \cite{Geng_TIN_opt}) so they will not be repeated here.

\section{TIM-TIN Problem Formulation}

Consider the general $K$-user interference channel presented in (\ref{equ_channel}). 
Suppose over $n$ channel uses each user $k\in\{1,2,...,K\}$ sends out $b_k$ independent scalar data streams, each of which carries one symbol $s_{k,l}$ and is transmitted along the $n\times 1$ beamforming vector $\mathbf{v}_{k,l}$, $l\in\{1,2,...,b_k\}$. Note the symbol $s_{k,l}$ comes from independent Gaussian codebooks, each with zero mean and unit power, and the beamforming vectors $\mathbf{v}_{k,l}$ are scaled to have unit norm. Then over $n$ channel uses, at receiver $k$ we obtain the $n\times 1$ vector,

\begin{equation}
\mathbf{y}_k=\sum_{i=1}^K\sum_{l=1}^{b_i}\sqrt{P^{\alpha_{ki}}}e^{j\theta_{ki}}\sqrt{P^{r_{i,l}}}\mathbf{v}_{i,l}s_{i,l}+\mathbf{z}_k
\end{equation}
where  $\mathbf{z}_k$ is the $n\times 1$ AWGN vector at receiver $k$, and $P^{r_{i,l}}$ is the power allocated to the $l$-th data stream of user $i$. All $r_{i,l}\leq 0$ because of the power constraint.

For receiver $k$, the covariance matrix of the desired signal is
\begin{equation}
\mathbf{Q}^D_k=\sum_{l=1}^{b_k}(\mathbf{v}_{k,l}\mathbf{v}_{k,l}^\dag)P^{r_{k,l}+\alpha_{kk}}
\end{equation}
and the covariance matrix of the interference from transmitter $i\neq k$
\begin{equation}
\mathbf{Q}_{ki}=\sum_{l=1}^{b_i}(\mathbf{v}_{i,l}\mathbf{v}_{i,l}^\dag)P^{r_{i,l}+\alpha_{ki}}
\end{equation}
so that the covariance matrix of the net interference-plus-noise
\begin{equation}
\mathbf{Q}^{N+I}_k=\sum_{i\neq k}\mathbf{Q}_{ki}+\mathbf{I}
\end{equation}
where $\mathbf{I}$ is an $n\times n$ identity matrix.

Then for each user $k$, given the beamforming vectors and power allocation for each data stream, the achievable rate per channel use is
\begin{equation}
\label{Rate_general}
\begin{aligned}
R_k&=\frac{1}{n}I(s_{k,1},s_{k,2},...,s_{k,b_k};\mathbf{y}_k)\\
&=\frac{1}{n}[h(\mathbf{y}_k)-h(\mathbf{y}_k|s_{k,1},s_{k,2},...,s_{k,b_k})]\\
&=\frac{1}{n}\big\{\log[\det(\mathbf{Q}^D_k+\mathbf{Q}^{N+I}_k)]-\log[\det(\mathbf{Q}^{N+I}_k)]\big\}
\end{aligned}
\end{equation}
and the GDoF $d_k$ can be written as
\begin{equation}
\label{GDoF_general}
d_k=\lim_{P\rightarrow \infty}\frac{R_k}{\log P}=\lim_{P\rightarrow \infty}\frac{\log[\det(\mathbf{Q}^D_k+\mathbf{Q}^{N+I}_k)]-\log[\det(\mathbf{Q}^{N+I}_k)]}{n\log P}
\end{equation}
To optimize a GDoF tuple for such a TIM-TIN scheme, it is clear that one needs to optimize over both the directions of all beamforming vectors and the allocated power for all data streams.

Let us simplify the achievable GDoF expression into a more intuitive form.  Consider a term of the type $\log[\det(\mathbf{I}+P^{\kappa_1}\mathbf{v}_1\mathbf{v}_1^\dag+P^{\kappa_2}\mathbf{v}_2\mathbf{v}_2^\dag+...+P^{\kappa_m}\mathbf{v}_m\mathbf{v}_m^\dag)]$,
where $\mathbf{v}_i$, $i\in\{1,2,...,m\}$ are $n\times 1$ beamforming vectors. We assume $\kappa_1\geq\kappa_2\geq...\geq\kappa_m\geq0$ without loss of generality. Then we consider the beamforming vectors one by one. For $\mathbf{v}_1$, we relabel it as $\mathbf{v}_{\Pi(1)}$ and correspondingly its associated power exponent $\kappa_{1}$ as $\kappa_{\Pi(1)}$. For $\mathbf{v}_2$, if it falls into $\mathrm{span}(\mathbf{v}_{\Pi(1)})$, we remove it and then proceed to $\mathbf{v}_{3}$; otherwise, we relabel it as $\mathbf{v}_{\Pi(2)}$ and correspondingly its associated power exponent $\kappa_{2}$ as $\kappa_{\Pi(2)}$. We repeat this operation for each beamforming vector. In words, for $\mathbf{v}_{i}$, if it falls into $\mathrm{span}(\mathbf{v}_{\Pi(1)}, \mathbf{v}_{\Pi(2)},...,\mathbf{v}_{\Pi(l)})$, which is the span of all previous linearly independent vectors we obtain from $\{\mathbf{v}_1, \mathbf{v}_2,...,\mathbf{v}_{i-1}\}$, we remove it and then proceed to $\mathbf{v}_{i+1}$; otherwise, we relabel it as $\mathbf{v}_{\Pi(l+1)}$ and correspondingly its associated power exponent $\kappa_{i}$ as $\kappa_{\Pi(l+1)}$. Finally, we will have $\gamma\leq n$ linearly independent beamforming vectors $\mathcal{V}_{\Pi} =\{\mathbf{v}_{\Pi(1)}, \mathbf{v}_{\Pi(2)},...,\mathbf{v}_{\Pi(\gamma)}\}$ and their associated power exponents $\mathcal{P}_{\Pi}=\{ \kappa_{\Pi(1)},\kappa_{\Pi(2)},...,\kappa_{\Pi(\gamma)}\}$. Based on the above definitions, we have the following lemma.

\begin{lemma}\label{Matrix_lemma}
Suppose $\mathbf{v}_i$, $i\in\{1,2,...,m\}$ are $n\times 1$ vectors, and $\kappa_1\geq\kappa_2\geq...\geq\kappa_m\geq0$, then
\begin{equation}
\begin{aligned}
&\log[\det(\mathbf{I}+P^{\kappa_1}\mathbf{v}_1\mathbf{v}_1^\dag+P^{\kappa_2}\mathbf{v}_2\mathbf{v}_2^\dag+...+P^{\kappa_m}\mathbf{v}_m\mathbf{v}_m^\dag)]\\
=&\sum_{i=1}^\gamma \kappa_{\Pi(i)}\log P +o(\log(P))\nonumber
\end{aligned}
\end{equation}
\end{lemma}

The proof is presented in the full paper \cite{Geng_Sun_Jafar}. Applying Lemma \ref{Matrix_lemma} to the two terms $\log[\det(\mathbf{Q}^D_k+\mathbf{Q}^{N+I}_k)]$ and $\log[\det(\mathbf{Q}^{N+I}_k)]$ in (\ref{GDoF_general})\footnote{If in (\ref{GDoF_general}), the receiver power exponents of certain data streams are less than $0$, we can ignore these streams without impacting the GDoF result.},  the  TIM-TIN problem is simplified into a form where the dependence on the assigned vector spaces and power levels is  explicit.


To understand the encoding/decoding scheme, note that according to the chain rule for the mutual information, (\ref{Rate_general}) can also be written as
\begin{equation}
\label{chain_rule}
R_k=\frac{1}{n}\sum_{i=1}^{b_k}I(s_{k,i};\mathbf{y}_k|s_{k,1},...,s_{k,i-1})
\end{equation}
From the right hand side of (\ref{chain_rule}), we can obtain the GDoF for each desired data stream $s_{k,i}$,
\begin{equation}
\label{GDoF_1beam}
\begin{aligned}
d_{k,i}&=\lim_{P\rightarrow \infty}\frac{I(s_{k,i};\mathbf{y}_k|s_{k,1},...,s_{k,i-1})}{n\log P}\\
&=\lim_{P\rightarrow \infty}\frac{h(\mathbf{y}_k|s_{k,1},...,s_{k,i-1})-h(\mathbf{y}_k|s_{k,1},...,s_{k,i-1},s_{k,i})}{n\log P}
\end{aligned}
\end{equation}
Applying Lemma \ref{Matrix_lemma} to (\ref{GDoF_1beam}), then summing up all the GDoFs for each desired data stream $d_{k,i}$ $i\in\{1,2,...,b_k\}$, we can obtain the same result for $d_k$ as in (\ref{GDoF_general}). Interesting, (\ref{chain_rule}) and (\ref{GDoF_1beam}) indicate that $d_k$ can be obtained in a successive cancellation manner. In other words, we can first decode $s_{k,1}$ from the received signal at receiver $k$, whose achievable rate is  $I(s_{k,1};\mathbf{y}_k)$. Then from (\ref{GDoF_1beam}), we can obtain the GDoF $d_{k,1}$ for data stream $s_{k,1}$. After decoding $s_{k,1}$, the receiver $k$ can subtract it from the received signal and then decode $s_{k,2}$, whose achievable rate is  $I(s_{k,2};\mathbf{y}_k|s_{k,1})$.  Similarly, from (\ref{GDoF_1beam}), we can obtain $d_{k,2}$. We can repeat this decode-and-subtract procedure to get the GDoFs of all desired data streams for user $k$, which lead to the final result $d_{k}$.

\begin{figure}[h]
\begin{center}
 \includegraphics[width= 7 cm]{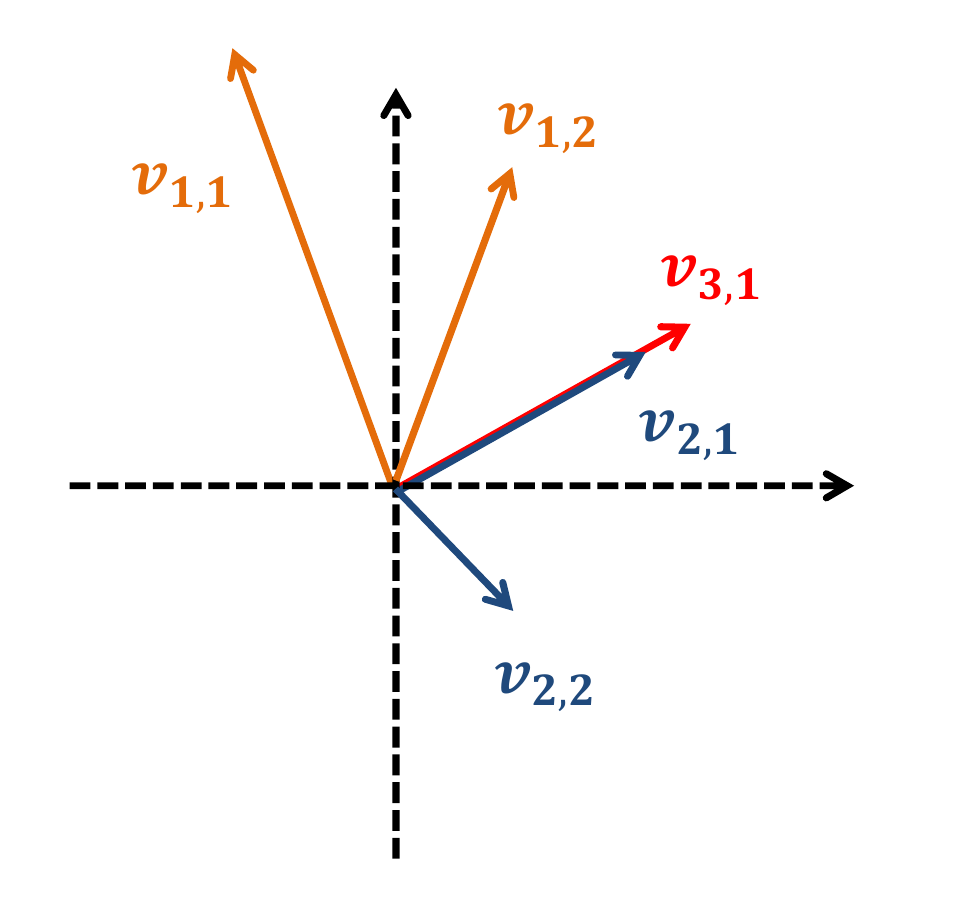}
 \caption{The received signal at receiver 1, where the length of the vector represents the received power of the carried symbol. }
\label{exp_GDoF}
\end{center}
\end{figure}

\begin{ex}
Consider a $3$-user interference channel, in which over $2$ channel uses the users $1$, $2$ and $3$ deliver $2$, $2$ and $1$ data streams, respectively. Suppose given the beamforming vectors, the transmitted power allocated to each symbol and channel strength levels for each link, the received signal at receiver $1$ is depicted in Fig.\ref{exp_GDoF}, in which $\mathbf{v}_{2,1}$ and $\mathbf{v}_{3,1}$ are aligned along one direction. The length of the vector represents the received power of the carried symbol, and
\begin{align*}
r_{1,1}+\alpha_{11}>r_{1,2}+\alpha_{11}>r_{3,1}+\alpha_{13}>r_{2,1}+\alpha_{12}>r_{2,2}+\alpha_{12}>0.
\end{align*}
Define
\begin{align}
d_k'&=\lim_{P\rightarrow \infty}\frac{\log[\det(\mathbf{Q}^D_k+\mathbf{Q}^{N+I}_k)]}{\log P}\\
d_k''&=\lim_{P\rightarrow \infty}\frac{\log[\det(\mathbf{Q}^{N+I}_k)]}{\log P}
\end{align}
We apply Lemma \ref{Matrix_lemma} to the above two terms, and after some manipulations, we have
\begin{align}
d_1'&= r_{1,1}+\alpha_{11}+r_{1,2}+\alpha_{11}\\
d_1''&=r_{3,1}+\alpha_{13}+r_{2,2}+\alpha_{12}
\end{align}
Then the GDoF of user $1$ is
\begin{equation}
\label{Exp_GDoF_equ}
\begin{aligned}
d_1&=\frac{d_1'-d_1''}{2}\\
&=\frac{1}{2}[(r_{1,1}+\alpha_{11}+r_{1,2}+\alpha_{11})-(r_{3,1}+\alpha_{13}+r_{2,2}+\alpha_{12})]
\end{aligned}
\end{equation}

In the following, we illustrate how to obtain the same GDoF result through successive cancellation. To decode $s_{1,1}$, we first zero-force the strongest interference $s_{1,2}$ and then treat all the other interference as noise. Then the GDoF of data stream $s_{1,1}$ is
\begin{equation}
\begin{aligned}
d_{1,1}&=\frac{(r_{1,1}+\alpha_{11}-\max\{r_{3,1}+\alpha_{13},r_{2,1}+\alpha_{12},r_{2,2}+\alpha_{12}\})}{2}\\
&=\frac{1}{2}(r_{1,1}+\alpha_{11}-r_{3,1}-\alpha_{13})
\end{aligned}
\end{equation}
After recovering $s_{1,1}$, we can subtract it off from the received signal and then decode $s_{1,2}$. Similarly, we still first zero-force the strongest interference $s_{3,1}$\footnote{Obviously, the aligned interference $s_{2,1}$ is also zero-forced simultaneously.} and then treat all the other interference as noise. The GDoF of data stream $s_{1,2}$ is
\begin{equation}
\begin{aligned}
d_{1,2}=\frac{1}{2}(r_{1,2}+\alpha_{11}-r_{2,2}-\alpha_{12})
\end{aligned}
\end{equation}
The GDoF for user $1$ is the sum of $d_{1,1}$ and $d_{1,2}$,
\begin{align}
d_1=\frac{1}{2}[(r_{1,1}+\alpha_{11}+r_{1,2}+\alpha_{11})-(r_{3,1}+\alpha_{13}+r_{2,2}+\alpha_{12})]
\end{align}
which equals  (\ref{Exp_GDoF_equ}).
\end{ex}

\section{A Baseline: TIM-TIN Decomposition}

Since the joint optimization of signal power levels and signal vector spaces for the general TIM-TIN problem seems challenging, we present a natural baseline based on a decomposition of the problem into TIM and TIN components, which can be solved separately and then combined to produce an achievable GDoF tuple for the original problem. 
A  TIM-TIN decomposition is defined as follows.  Given an interference network, two copies of the network are created, called the TIM component and the TIN component. The desired links are copied in both networks. However, each interfering link is mapped to either the TIM component or the TIN component (but not both). Note that  many different TIM-TIN decompositions are possible. 

\begin{figure}[!t]
\centering
\subfigure[]{
\includegraphics[width= 4 cm]{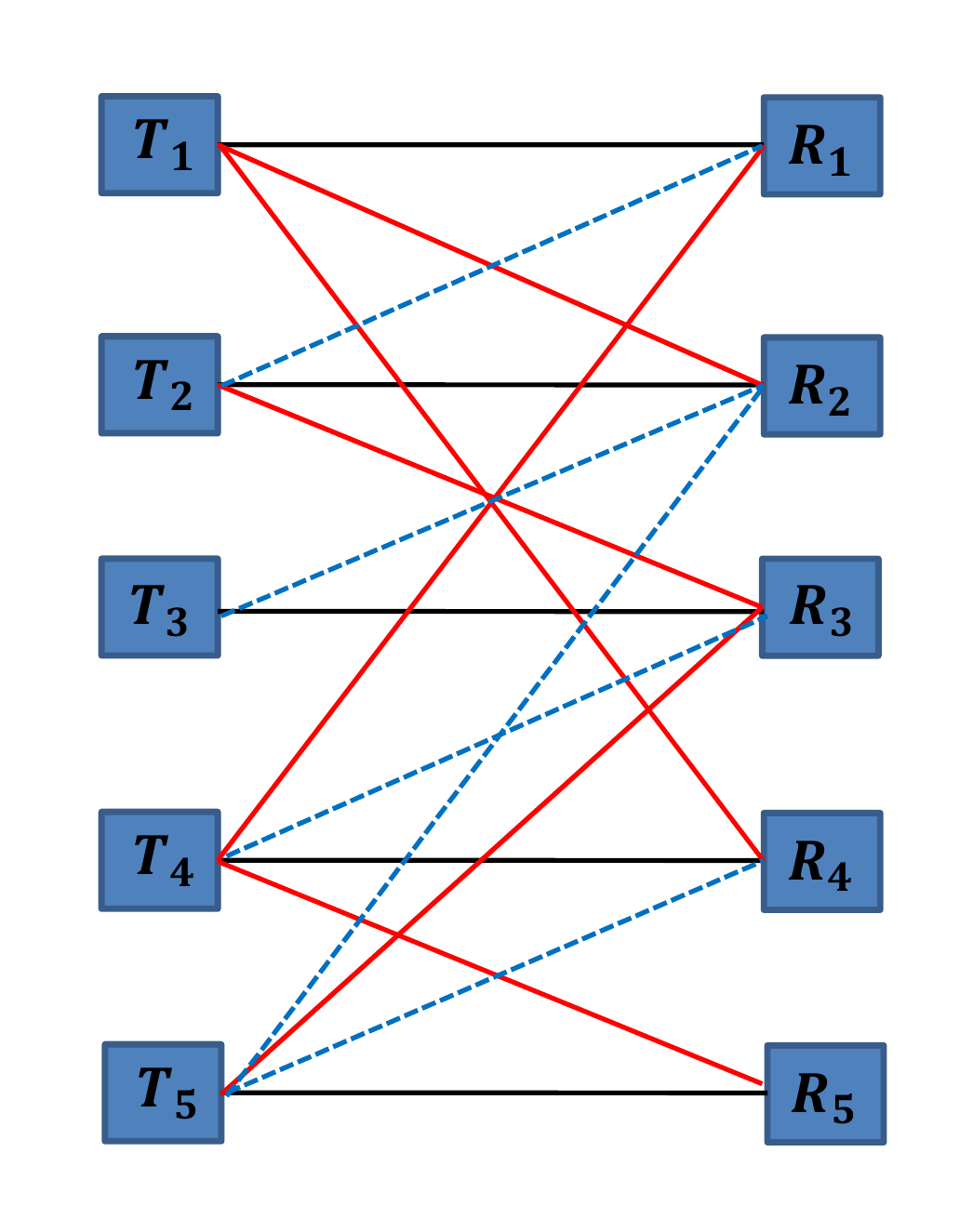}
\label{5user}}\\
\subfigure[]{
\includegraphics[width= 4 cm]{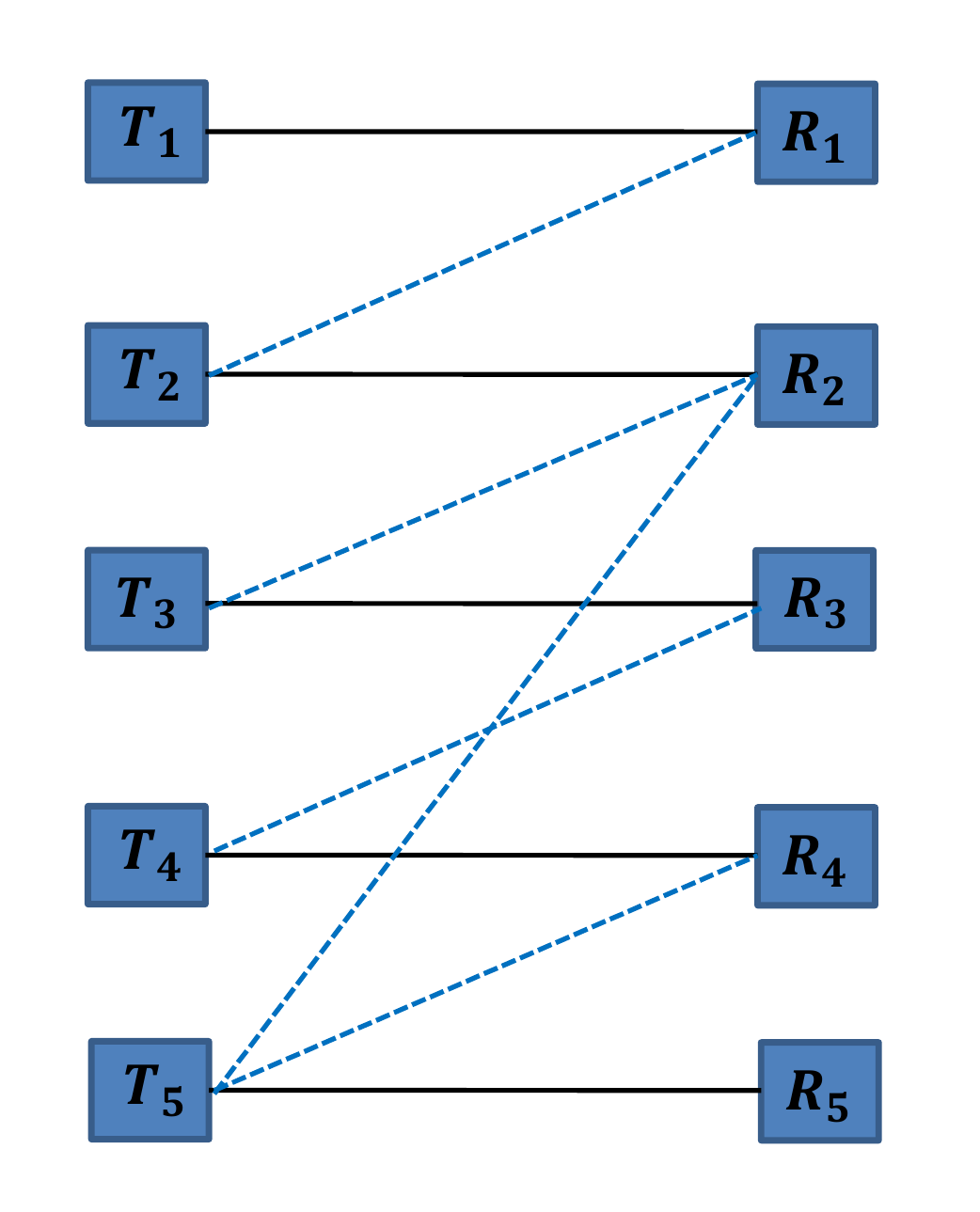}
\label{5user_M1}}
\subfigure[]{
\includegraphics[width= 4 cm]{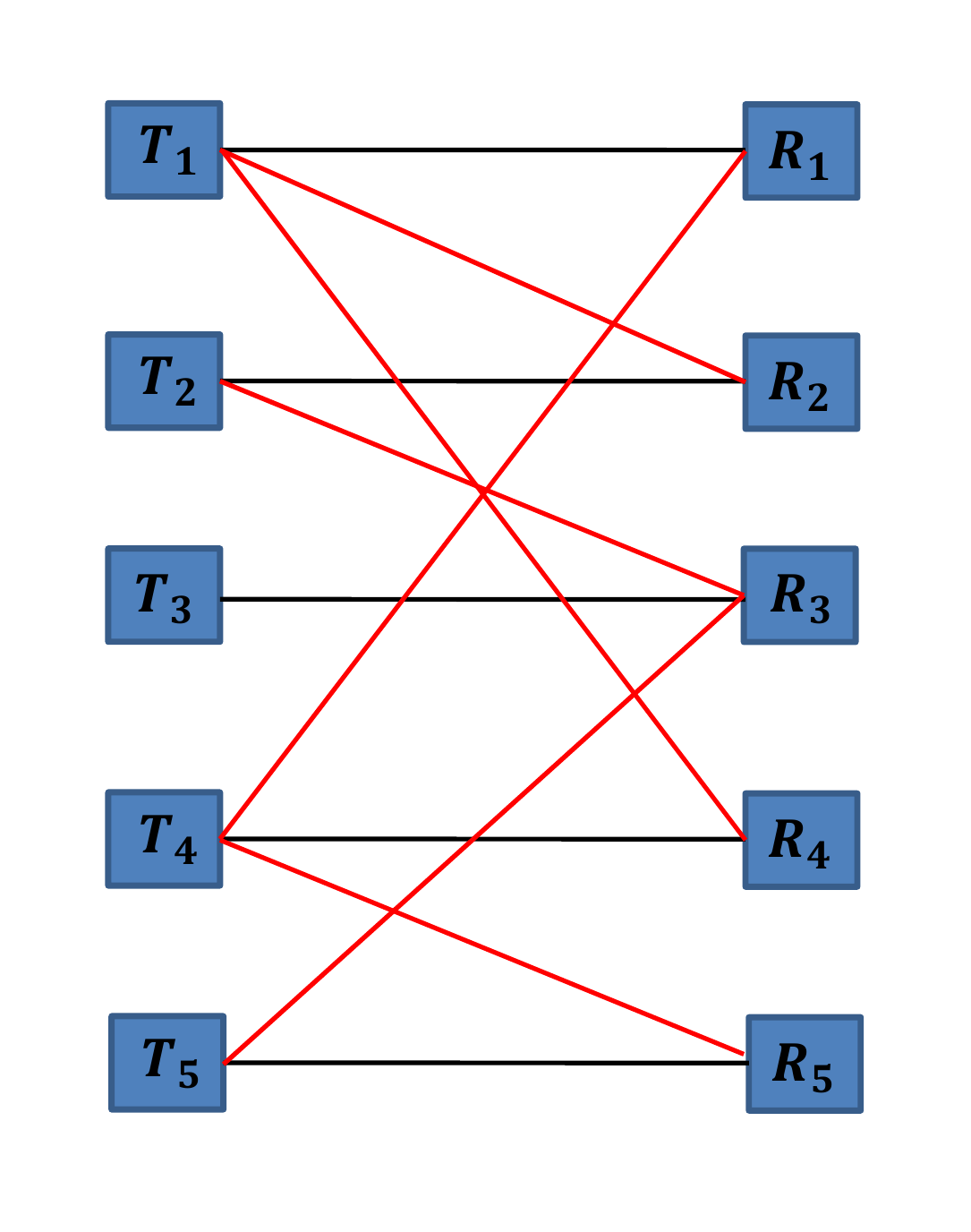}
\label{5user_S1}}
\subfigure[]{
\includegraphics[width= 3.7 cm]{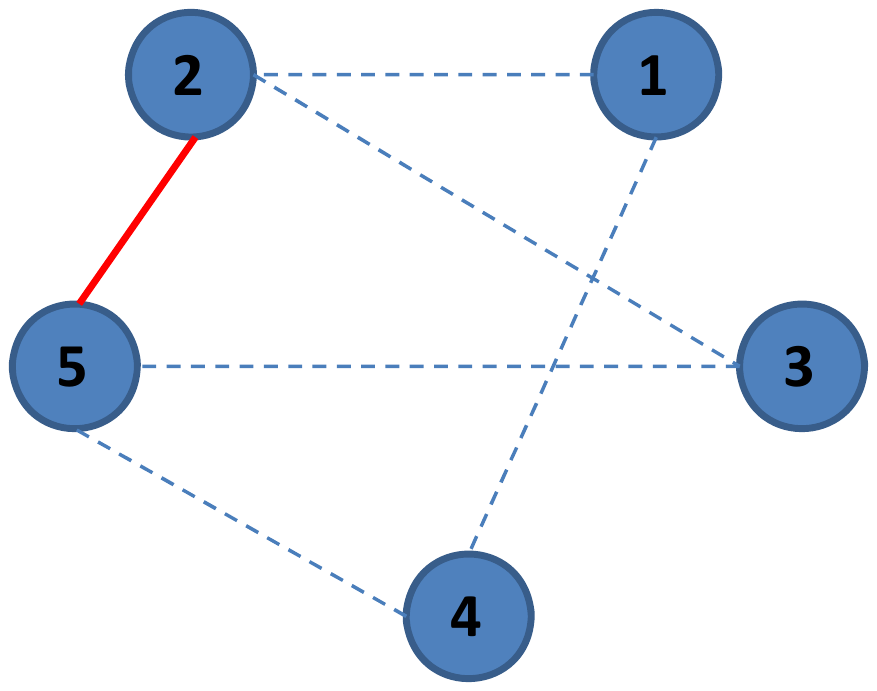}
\label{5user_align1}}
\subfigure[]{
\includegraphics[width= 3.7 cm]{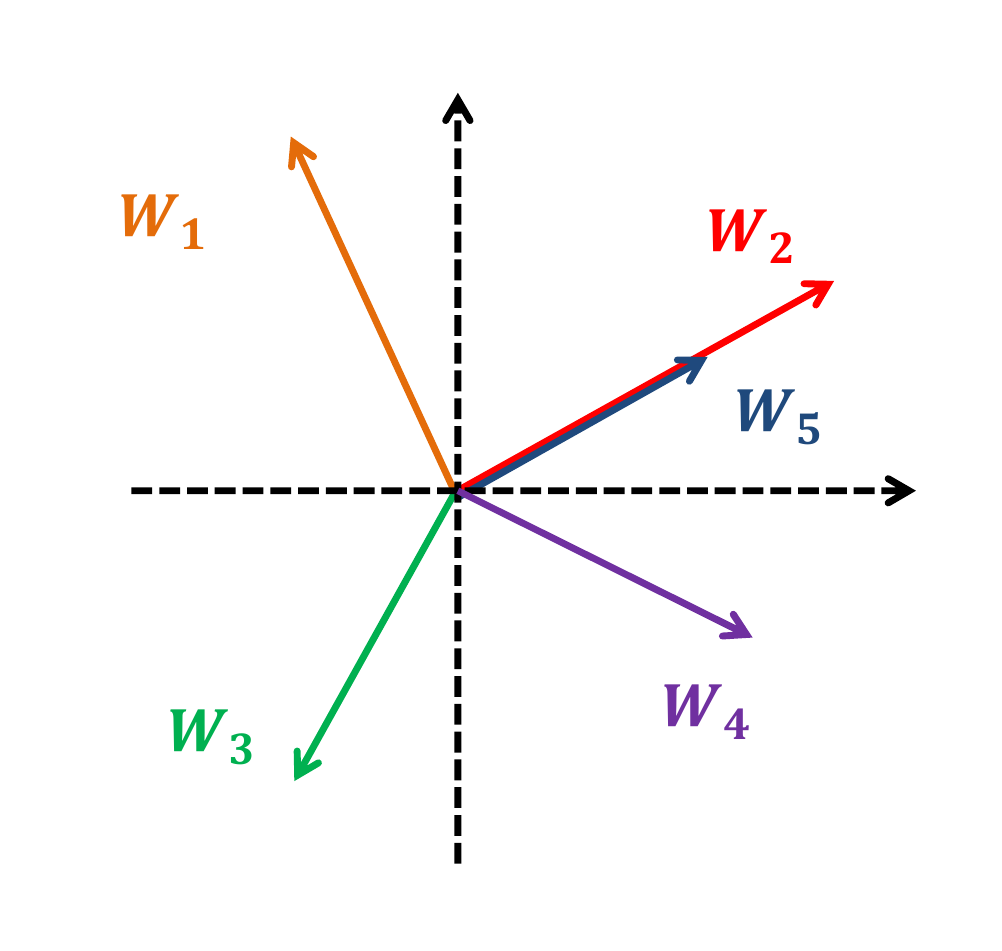}
\label{sol2}}
\caption[]{
\subref{5user} 5-user interference channel; \subref{5user_M1} The TIN component ; \subref{5user_S1} The TIM component; \subref{5user_align1} The alignment graph (red solid lines) for the TIM component showing conflicts as dashed blue lines; \subref{sol2} An achievable scheme to achieve  symmetric GDoF of 0.3.}
\label{5user_1}
\end{figure}

In Fig. \ref{5user_1}, we give an example to show one possible decomposition of a $5$-user interference channel into a TIN component and a TIM component. In this figure, the black and red links have strength 1.0 and the blue links have strength 0.5. In the TIM-TIN decomposition shown in the figure, the blue interference links are mapped to the TIN component and the red ones to the TIM component.

The purpose of the TIM-TIN decomposition is to simplify the problem by solving the TIM and TIN components separately. First consider the TIM component only. We assume all the non-zero links that are mapped to the TIM component are equally strong (even if they are not) and find a linear TIM solution to obtain the GDoF tuple $(d_{1,\mathrm{TIM}},d_{2,\mathrm{TIM}},...,d_{K,\mathrm{TIM}})$, which identifies the \emph{fraction} of the interference-free signal space that is available to each user.  Next, we consider the TIN component only. Suppose through appropriate power control and TIN, the GDoF tuple $(d_{1,\mathrm{TIN}},d_{2,\mathrm{TIN}},...,d_{K,\mathrm{TIN}})$ is achievable, which identifies  the available signal power levels. Then the product of the two  fractions for each user, i.e., the GDoF tuple $(d_{1,\mathrm{TIN}}\times d_{1,\mathrm{TIM}},d_{2,\mathrm{TIN}}\times d_{2,\mathrm{TIM}},...,d_{K,\mathrm{TIN}}\times d_{K,\mathrm{TIM}})$ is achievable, which identifies the net  signal dimensions available to each user by this decomposition-based approach. The convex hull of all similarly achieved GDoF-tuples corresponding to different TIM-TIN decompositions is also achievable through time-sharing.

In Fig. \ref{5user_1}(c), the TIM component achieves symmetric DoF of $d_{sym,TIM}=0.5$. This is seen from the alignment and conflict graphs shown in Fig. \ref{5user_1}(d) because there are no internal conflicts \cite{Jafar_TIM}. In the TIN component, which contains all the  interfering links with strength 0.5, its symmetric GDoF is determined by the longest zigzag chain embedded in the network topology \cite{Geng_TIN_opt}. Since the length of the longest zigzag chain is $l^*=4$, the symmetric GDoF of this network is $d_{sym,TIN}=\frac{1+0.5l^*}{1+l^*}=0.6$. Therefore, by TIM-TIN decomposition the achievable symmetric GDoF for the original network is $0.6\times0.5=0.3$. The achievable scheme is shown explicitly in Fig. \ref{5user_1}\subref{sol2}. It uses a 2 dimensional space and 4 beamforming vectors, where any two of them are linearly independent and $W_2$ and $W_5$ are aligned along the same vector. The transmit powers are selected as $P_1=P^0=1$, $P_2=P^{-0.1}$, $P_3=P^{-0.2}$, $P_4=P^{-0.3}$ and $P_5=P^{-0.4}$. It is easy to verify that by using this achievable scheme, every user can achieve the GDoF of $0.3$.
\begin{itemize}
\item Receiver 1 first zero-forces the interference from transmitter 4 (denoted as $I_4$). Then, in the remaining signal dimensions, it treats the interference $I_2$ as noise. Therefore, the achievable GDoF for receiver $1$ is $(1-0.4)/2=0.3$.

\item Receiver 2 first zero-forces $I_1$ and then treats $I_3$ and $I_5$ as noise to get $(0.9-0.3)/2=0.3$ GDoF.

\item Receiver 3 first zero-forces $I_2$ and $I_5$, and then treats $I_4$ as noise to get $(0.8-0.2)/2=0.3$ GDoF.

\item Receiver 4 first zero-forces $I_1$ and then treats $I_5$ as noise to get $(0.7-0.1)/2=0.3$ GDoF.

\item Receiver 5 only zero-forces $I_4$ to get $0.6/2=0.3$ GDoF.
\end{itemize}

\begin{figure}[h]
\centering
\subfigure[]{
\includegraphics[width= 4 cm]{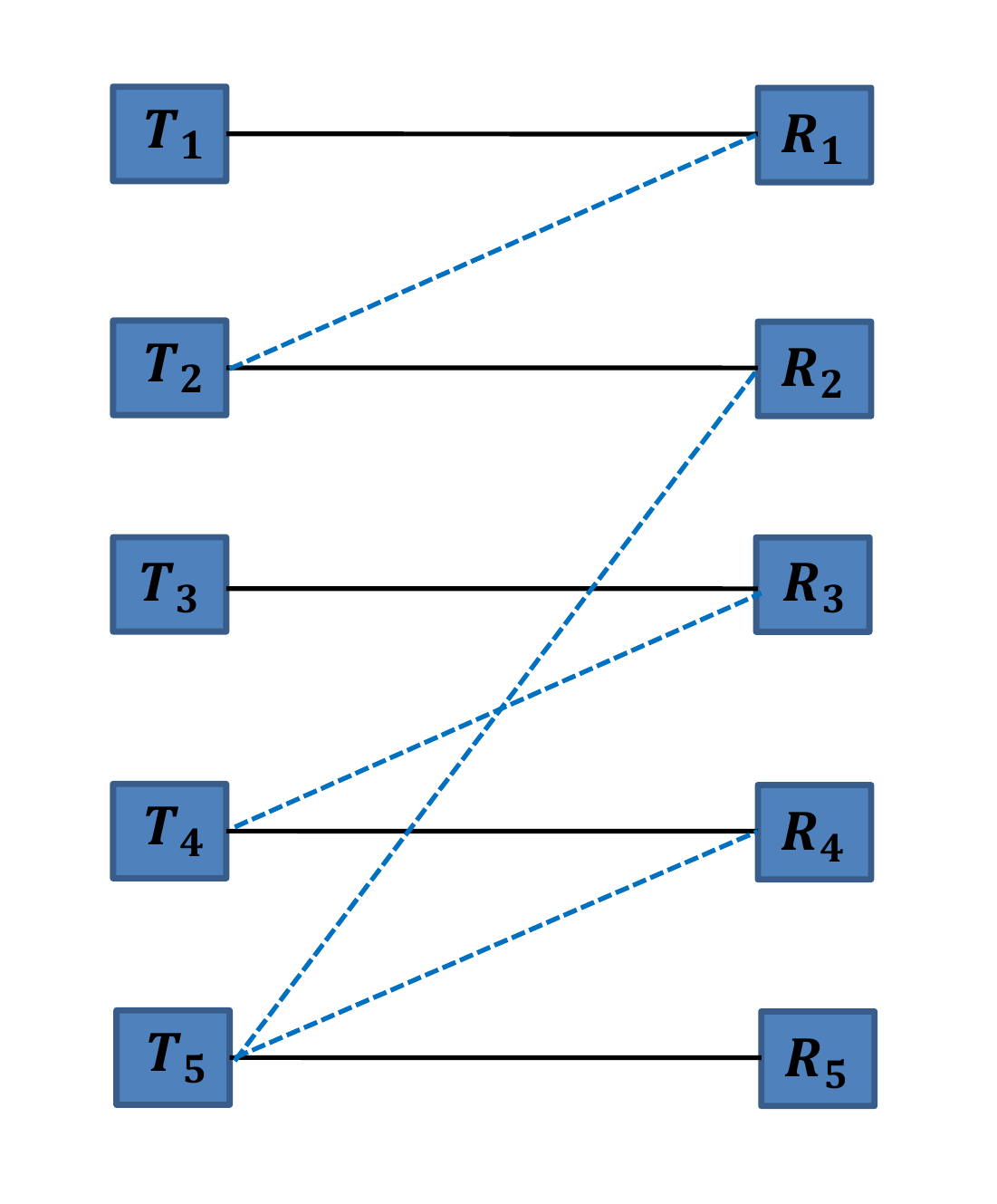}
\label{5user_M2}}
\subfigure[]{
\includegraphics[width= 4 cm]{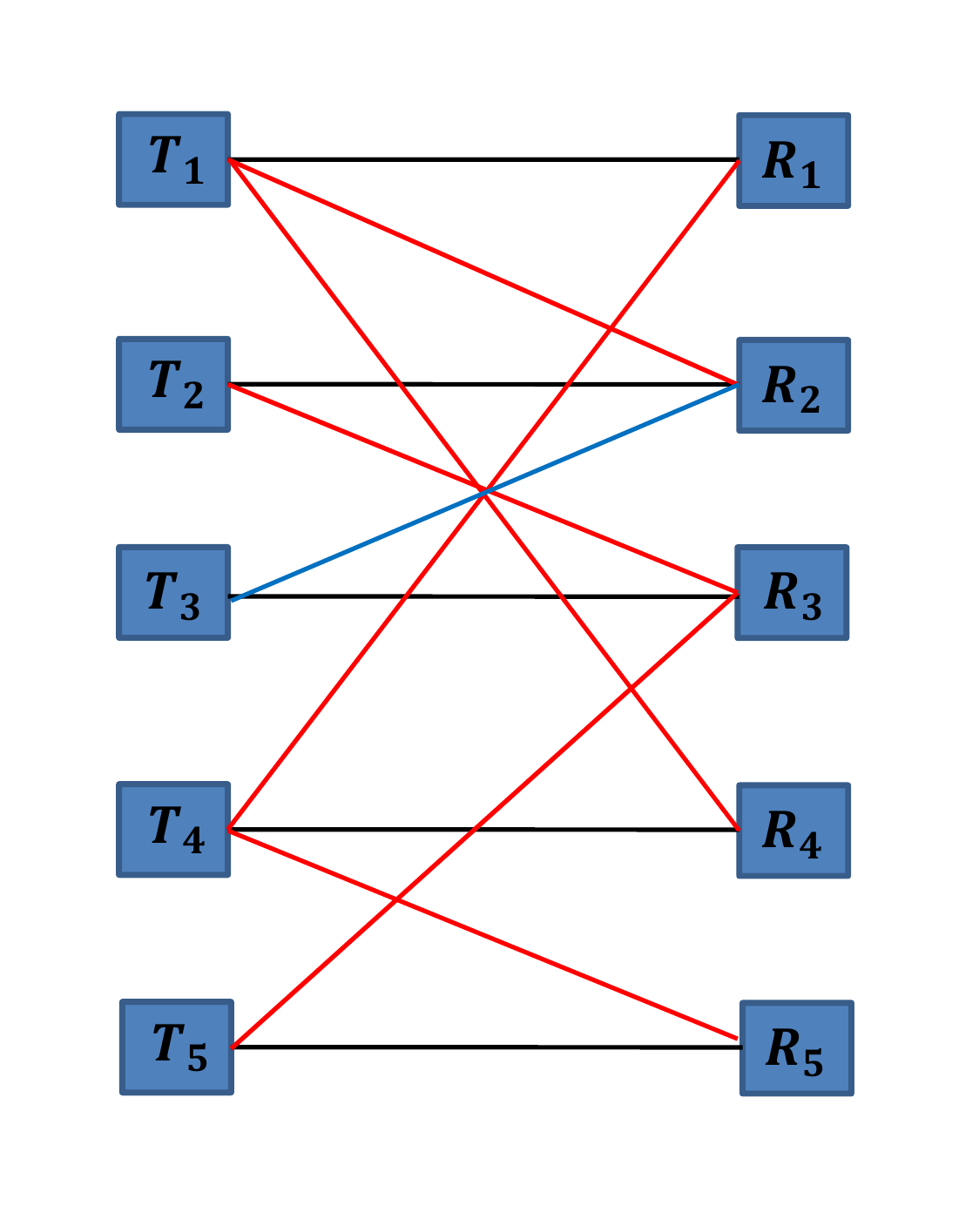}
\label{5user_S2}}
\subfigure[]{
\includegraphics[width= 4 cm]{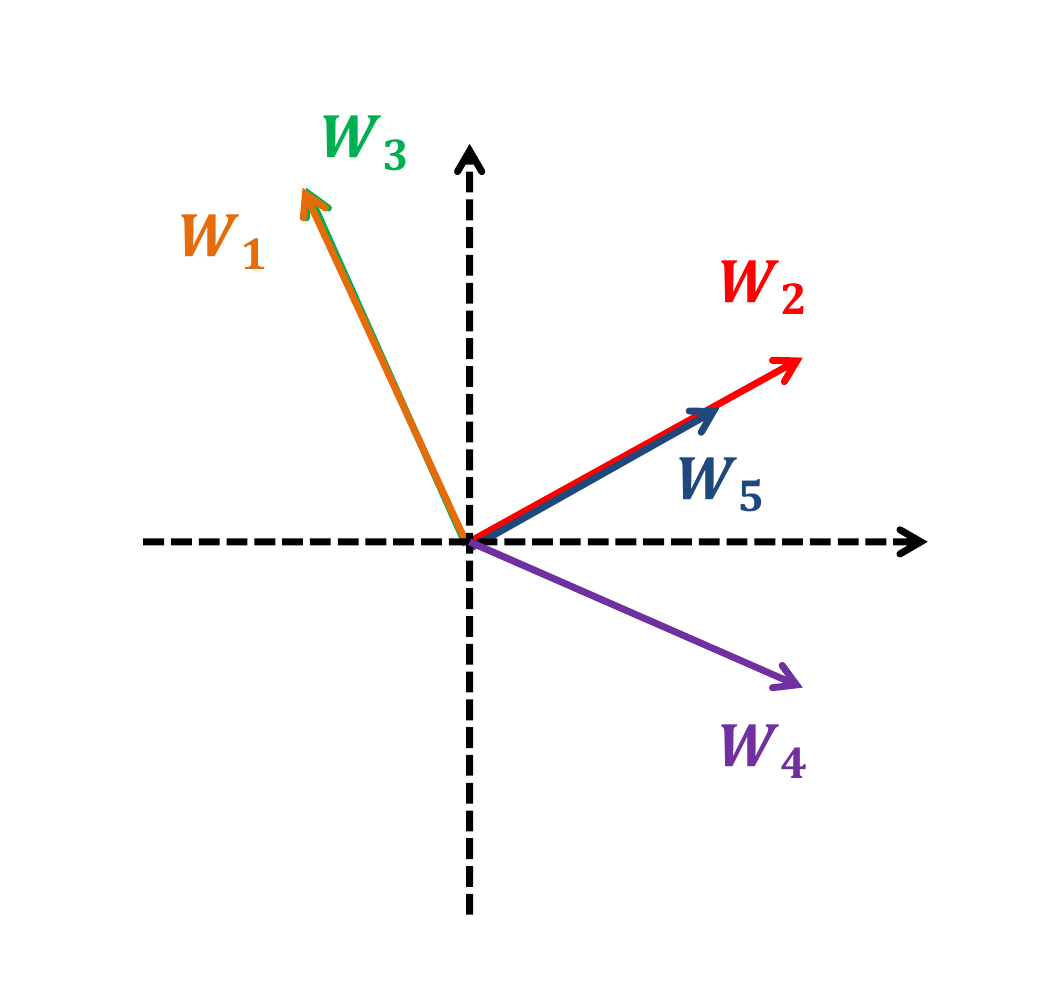}
\label{sol3}}
\caption[]{
\subref{5user_M2} The TIN component  whose symmetric GDoF is $\frac{2}{3}$; \subref{5user_S2} The TIM component  whose symmetric GDoF is $\frac{1}{2}$. In this figure, we treat the medium interfering link $h_{23}$ (the blue solid line) as a strong interfering link; \subref{sol3} The achievable scheme to achieve symmetric GDoF of $\frac{1}{3}$.}
\label{5user_2}
\end{figure}

While the TIM-TIN decomposition is  flexible, i.e., any interfering link can be mapped into either TIM or TIN components, generally one would expect that to obtain a ``good'' achievable GDoF region,  the TIM component should contain the appropriate ``strong'' interfering links and the TIN component should contain the appropriate ``weak'' interfering links. This is, however, not always the case.

For the example of Fig. \ref{5user_1}, it is not difficult to verify that if we move the medium interfering link $l_{23}$ between transmitter $3$ and receiver $2$ from the TIN component to the TIM component, then the symmetric GDoF for the new TIN component becomes $\frac{2}{3}$ and the symmetric GDoF for the new TIM component remains as $\frac{1}{2}$. Therefore, the achievable symmetric GDoF via TIM-TIN decomposition can be improved  to $\frac{1}{3}$. The corresponding TIN component, TIM component and achievable scheme are shown in Fig. \ref{5user_2}. The achievable scheme uses a 2-dimensional vector space and 3 beamforming vectors, any two of which are linearly independent. $W_1$ and $W_3$ are aligned along one direction, and $W_2$ and $W_5$ are aligned along another direction. The transmit powers are $P_1=P_3=P^0=1$, $P_2=P_4=P^{-\frac{1}{6}}$, and $P_5=P^{-\frac{1}{3}}$.

\section{Conclusion}
We formulated a signal vector space and signal power level allocation problem (the TIM-TIN problem) under the assumption that only a coarse knowledge of channel strengths and no knowledge of channel phases is available to the transmitters. A natural decomposition of the problem into TIN and TIM components was proposed as a baseline. Applications of TIM-TIN decomposition to interesting network topologies are currently being explored and will be reported in future work.


\begin{thebibliography}{10}
\providecommand{\url}[1]{#1} \csname url@rmstyle\endcsname
\providecommand{\newblock}{\relax} \providecommand{\bibinfo}[2]{#2}
\providecommand\BIBentrySTDinterwordspacing{\spaceskip=0pt\relax}
\providecommand\BIBentryALTinterwordstretchfactor{4}
\providecommand\BIBentryALTinterwordspacing{\spaceskip=\fontdimen2\font
plus \BIBentryALTinterwordstretchfactor\fontdimen3\font minus
  \fontdimen4\font\relax}
\providecommand\BIBforeignlanguage[2]{{%
\expandafter\ifx\csname l@#1\endcsname\relax
\typeout{** WARNING: IEEEtran.bst: No hyphenation pattern has been}%
\typeout{** loaded for the language `#1'. Using the pattern for}%
\typeout{** the default language instead.}%
\else \language=\csname l@#1\endcsname \fi #2}}


%


\bibitem{Jafar_FnT}
S.~Jafar, ``Interference Alignment: A new look at signal dimensions in a communication network'', Foundations and Trends in Communications and Information Theory, vol.7, no.1, pp. 1-136.

\bibitem{Jafar_TIM}
S.~Jafar, ``Topological interference management through index coding," e-print arXiv:1301.3106



\bibitem{Annapureddy_Veeravalli_TIN_opt}
V.~Annapureddy and V.~Veeravalli, ``Gaussian interference networks: sum capacity in the low-interference regime and new outer bounds on the capacity region," \textit{IEEE Transactions on Information Theory}, vol. 55, no. 7, pp. 3032-3050, July, 2009.

\bibitem{Motahari_Khandani_TIN_opt}
A.~Motahari, and A.~Khandani, ``Capacity bounds for the Gaussain interference channel," \textit{IEEE Transactions on Inoformation Theory}, vol. 55, no. 2, pp. 620-643, Feb. 2009.

\bibitem{Kramer_TIN_opt}
X.~Shang, G.~Kramer, and B.~Chen, ``A new outer bound and the noisy-interference sum-rate capacity for Gaussian interference channels," \textit{IEEE Transactions on Information Theory}, vol. 55, no. 2, pp. 689-699, Feb. 2009.

\bibitem{Annapureddy_Veeravalli_MIMO}
V.~Annapureddy and V.~Veeravalli, ``Sum capacity of MIMO interference channels in the low interference regime," \textit{IEEE Transactions on Information Theory}, vol. 57, no. 5, pp. 2565-2581, May 2011.

\bibitem{Tse_GDoF}
R.~Etkin, D.~Tse, and H.~Wang, ``Gaussian interference channel capacity to within one bit,'' \textit{IEEE Transactions on Inforamtion Thoery}, vol. 54, no. 12, pp. 5534-5562, Dec. 2008.

\bibitem{Jafar_Vishwanath_GDoF}
S.~Jafar, and S.~Vishwanath, ``Generalized degrees of freedom of the symmetric Gaussian K user interference channel", \textit{IEEE Transactions on Information Theory}, vol. 56, no. 7, pp. 3297-3303, July 2010.

\bibitem{Geng_TIN_opt}
C.~Geng, N.~Naderializadeh, A.~Avestimeher, and S.~Jafar, ``On the optimality of treating interference as noise,'' e-print ArXiv:1305.4610.

\bibitem{Birk_Kol}
Z.~Bar-Yossef, Y.~Birk,, T.~Jayram, and T.~Kol, ``Index coding with side information,'' \textit{IEEE Transactions on Information Thoery}, vol.57, no.3, pp.1479-1494, Mar. 2011.

\bibitem{Maleki_Cadambe_Jafar_index}
H.~Maleki, V.~Cadambe, and S.~Jafar, ``Index coding -- an interference alignment perspective,'' e-print arXiv:1205.1483.

\bibitem{Geng_Sun_Jafar}
C. ~Geng, H. ~Sun, and S. ~Jafar, ``Multilevel Topological Interference Management", Full paper in preparation.


%
%
%
%







%
%
%
%
%
%
%
%
%
%


\end{thebibliography}
\end{document}